\documentclass[a4paper,11pt]{article}
\pdfoutput=1 

\usepackage{jcappub} 

\title{\boldmath Strongest constraint in $f(R) = R+ \alpha R^2$ gravity: stellar stability}

\author[a]{Juan M. Z. Pretel, }
\author[a]{Sergio E. Jor\'as}
\author[a,b]{and Ribamar R. R. Reis}

\affiliation[a]{Instituto de F\'\i sica, Universidade Federal do Rio de Janeiro, \\ CEP 21941-972 Rio de Janeiro, RJ, Brazil}
\affiliation[b]{Observat\'orio do Valongo, Universidade Federal do Rio de Janeiro, \\ CEP 20080-090 Rio de Janeiro, RJ, Brazil}

\emailAdd{juanzarate@if.ufrj.br}
\emailAdd{joras@if.ufrj.br}
\emailAdd{ribamar@if.ufrj.br}

\abstract{ In the metric approach of $f(R)$ theories of gravity, the fourth-order field equations are often recast as effective Einstein equations in the presence of standard matter and a curvature fluid (which gathers all the extra terms), always in the Jordan frame. In this picture, we investigate the strong gravity regime of the $f(R) = R+ \alpha R^2$ model. In particular, we focus on the stability of a compact star composed by a mixture of ordinary matter --- described by a polytropic equation of state --- and an effective curvature fluid in an otherwise standard Einstein gravity, so that we are able to apply the usual equations that govern the radial adiabatic oscillations of relativistic stars. Our new restriction on the free parameter is $\alpha \lesssim 2.4 \times 10^8\ \text{cm}^2$ in order to guarantee stellar stability, about $100$ times more restrictive than previous results (based on mass-radius relations alone) in the literature. }

\begin{document}
\maketitle
\flushbottom

\section{Introduction}

Although $f(R)$ theories have been studied in order to explain the accelerated expansion of the universe at cosmological scales (both in the early and late eras), such class of theories must also be tested at smaller scales, that is, at the astrophysical level. Therefore, it is important to investigate the physical characteristics of compact stars within the framework of $f(R)$ theories of gravity. One of the simplest modifications of General Relativity (GR from now on) is the theory popularly known as the Starobinsky model \cite{Starobinsky}, given by $f(R) = R+ \alpha R^2$, where the presence of the $R^2$ term can give rise to cosmic inflation in the early Universe. Although this model has been already discarded on cosmological grounds \cite{PhysRevD.75.083504} because it is not viable for late-time cosmic acceleration, such a theory has been widely used in the strong-gravity regime for the construction of compact objects (see references below) in order to probe the outcome of such high-curvature modifications and ultimately providing some insight on viable modifications.

The structure of compact stars is usually studied through two approaches: perturbative methods (where  $f(R)$ is considered a small perturbation from GR) \cite{Cooney, Arapoglu, Orellana, AstashenokCapOdi1, Alavirad, AstashenokCapOdi2} and non-perturbative ones (where the full fourth-order differential equations have to be solved) \cite{Yazadjiev1, PhysRevD.89.064019, AstashenokCapOdi3, AstashenokOdinDom}. In both of them, the gravitational mass (defined on the surface of the star) decreases with $\alpha$. Nevertheless, in the non-perturbative approach a ``gravitational sphere'' emerges outside the star, so that the astrophysical mass (measured by distant observers) of compact stars increases with $\alpha$ --- see Ref.~\cite{Fulvio, Fulvio2} for a discussion on the many definitions of mass in $R^2$ gravity. A comprehensive analysis about stellar structure models within the context of modified theories of gravity formulated in both metric and  metric-affine formalisms has recently been carried out in Ref. \cite{Olmo2020}.

The aforementioned works successfully construct mass-radius diagrams for both quark and neutron stars in equilibrium with a bonus: they allow star masses $M$ above the standard GR limit. Although the plain determination of the star (maximum) mass could, in principle, lead to a maximum value for $\alpha$, it does not yield a lower bound to $\alpha$, since the mass values coincide with the ones predicted by GR when $\alpha\to 0$. 

In spite of being a crucial point, the stability of such stars has not been studied yet in $f(R)$ gravity. In GR, a necessary condition for stellar stability along the sequence of equilibrium configurations is $d M/d\epsilon_{\rm cent} >0$ (where $\epsilon_{\rm cent}$ is the central energy density). However, such requirement cannot be naively applied here: $f(R)$ theories present an extra degree of freedom, which would, in principle, allow a new decay branch --- the emission of scalar gravitational waves from a radially oscillating star (which simply does not happen in GR). 

On the other hand, we know that a sufficient condition for stability of any physical system is that its lowest normal-mode frequency must be real \cite{Glendenning}. In the present paper we  apply the latter (and more general) requirement in $f(R)$ gravity, writing the extra terms in the modified Einstein equations as an effective energy-momentum tensor (also known as curvature fluid). The advantage of writing the field equations as effective Einstein equations is that the complexity of the problem in modified gravity is reduced to problems already known and well studied in GR. Then, the energy-momentum tensor in the effective Einstein equations $T_{\mu\nu}$ includes both fluid components and, therefore, we can use the traditional equations for radial oscillations of adiabatic relativistic stars in GR in order to study the stellar stability in $f(R)$ gravity. Similar adoptions for two-fluid stars were also considered in \cite{Ciarcelluti, PanotopoulosLopes, LopesPanotopoulos}.

The stability of non-rotating stars is controlled by normal modes in which the perturbations in the fluid are purely radial \cite{Chandrasekhar1, Chandrasekhar, Chanmugam, KokkotasRuoff, VathChanmugam, Gondek, PanotopoulosLopes2017}, while the theory of non-radial oscillations is used to analyze the emission of gravitational radiation from non-stationary sources \cite{ChandrasekharFerrari}. In fact, the radial perturbations in GR do not couple to gravitational radiation. According to Ref. \cite{Ferrari2008}, quasi-normal modes are the proper modes at which a compact object oscillates when is excited by a non-radial perturbation. They are said quasi-normal because they are damped by the emission of gravitational waves. In this respect, some important contributions have been made such as the study of quasinormal modes of compact stars in GR \cite{1967ApJThorne, PhysRevD.43.1768, PhysRevD.58.124012, Kokkotas1999} and recently in $R^2$ gravity by taking advantage of its mathematical equivalence with scalar-tensor theories \cite{PhysRevD.92.043009, PhysRevD.98.104047, PhysRevD.98.044032, BlazquezSalcedo2019}. However, the normal modes of radial oscillations in $f(R)$ gravity have not been calculated yet. In the present work we are interested in studying the stellar stability against radial pulsations in the Starobinsky model using a two-fluid formalism.

In our analysis we will consider compact stars with a polytropic Equation of State (EoS) for the standard matter, this is, $p^{\text{m}} = \bar{\kappa} (\rho^{\text{m}})^{1 + 1/n}$, with polytropic index $n = 1$ and $\bar{\kappa} = 100\ \text{km}^2$, in geometric units,  which are typical values to describe neutron stars \cite{KokkotasRuoff}. We decide to use this EoS for its analytical simplicity and because it is widely used in the literature to study self-gravitating objects \cite{PhysRevLett.77.4134, 10.1093/mnras/289.1.117, PhysRevD.58.124012, PhysRevLett.120.201104}. Despite its simplicity, the qualitative relation between the pressure and the energy density --- similar to those in more realistic EoS --- suffices to achieve a proof of concept:  to demonstrate the constraint power of stability under radial oscillations in the presence of two fluids.

As we will see further below, our approach yields the strongest constraints in $\alpha$ to the best of our knowledge. This work is organized as follows: In Section \ref{Sec2} we present the modified TOV equations and boundary conditions for $f(R) = R + \alpha R^2$ gravity. Section \ref{Sec3} is devoted to describe the higher-order curvature invariant in the Einstein-Hilbert action as an effective fluid. In Section \ref{Sec4} we briefly summarize the adiabatic radial oscillations in GR. Numerical solutions of stellar structure equations and discussion are reported in Section \ref{Sec5}. The paper ends up with our conclusions in Section \ref{Sec6}. Here we will denote $R$ as the Ricci scalar, $r_{\rm surf}$ as the radius of the star where the matter pressure vanishes, and we will adopt the signature $(-, +, + ,+)$.


\section{Modified TOV equations} \label{Sec2}

In $f(R)$ theories of gravity, the Einstein-Hilbert action is modified by a generic function of the Ricci scalar $R$. The Jordan frame \footnote{We shall keep the analysis in the Jordan frame throughout this work.}  action is given by
\begin{equation}\label{1}
    S = \frac{1}{2\kappa}\int d^4x\sqrt{-g}f(R) + S_{\text{m}} ,
\end{equation}
where $\kappa = 8\pi G/c^4$, $g$ is the determinant of the metric tensor $g_{\mu\nu}$, and $S_{\text{m}}$ denotes the action of ordinary matter. 

The metric formalism \cite{SotiriouFaraoni, Felice} --- which we follow in this work --- consists of varying the action (\ref{1}) with respect to the metric tensor and yields the modified Einstein equations:
\begin{equation}\label{2}
f_R R_{\mu\nu} - \dfrac{1}{2}g_{\mu\nu}f - \nabla_\mu\nabla_\nu f_R + g_{\mu\nu}\square f_R = \kappa T_{\mu\nu}^{\text{m}} ,
\end{equation}
where $T_{\mu\nu}^{\text{m}}$ is the energy-momentum tensor for ordinary  matter, $f_R(R) \equiv  df(R)/dR$, $\nabla_\mu$ is the covariant derivative, and $\square \equiv \nabla_\mu \nabla^\mu$ is the d'Alembert operator in the curved spacetime. The trace of the above equation determines the dynamics of $R$ for a given matter source, namely,
\begin{equation}\label{3}
    3\square f_R(R) + Rf_R(R) - 2f(R) = \kappa T^{\text{m}} . 
\end{equation}

For the background, we consider a static and spherically symmetric system whose spacetime is described by the usual line element
\begin{equation}\label{4}
    ds^2 = -e^{2\psi(r)}(dx^0)^2 + e^{2\lambda(r)}dr^2 + r^2d\Omega^2 ,
\end{equation}
where $x^0 = ct$, and $d\Omega^2 = d\theta^2 + \sin^2\theta d\phi^2$ is the line element on the unit 2-sphere. The standard matter source under consideration is described as a perfect fluid, for which the energy-momentum tensor is $T_{\mu\nu}^{\text{m}} = ( \epsilon^{\text{m}} + p^{\text{m}} )u_\mu u_\nu + p^{\text{m}}g_{\mu\nu}$, being $u_\mu$ the four-velocity of the fluid, $\epsilon^{\text{m}} = c^2\rho^{\text{m}}$ the energy density (where $\rho^{\text{m}}$ indicates mass density), and $p^{\text{m}}$ is the pressure. In GR, where $f(R) = R$, the Klein-Gordon-like equation (\ref{3}) is reduced to $R = -\kappa (3p^{\text{m}} - \epsilon^{\text{m}})$, i.e, an algebraic equation. In fact, in $f(R)$ gravity, $T^{\text{m}} =0$ no longer implies $R = 0$ as in the outer region of a compact star in GR. Now the Ricci scalar (besides the metric itself) is also a dynamical field described by a differential equation (\ref{3}) --- this is the aforementioned extra scalar degree of freedom.

The theory of gravity to be used is the Starobinsky model $f(R) = R + \alpha R^2$ ---  also known as {\it quadratic gravity} --- where $\alpha$ is a free parameter which is usually given in units of $r_g^2$, where $r_g = GM_\odot/c^2 \approx 1.477\ \text{km}$ is the solar mass in geometrical units. Such theory has to be free of tachyonic instabilities which requires that $f_{RR} \equiv d^2f(R)/dR^2 \geq 0$ \cite{SotiriouFaraoni, Felice}, and hence $\alpha$ must be a positive constant. Through the geodetic precession of a gyroscope in a gravitational field and the precession of binary pulsars, this parameter has been constrained from the result of the Gravity Probe B experiment as $\alpha \lesssim 5 \times 10^{15}\  \text{cm}^2 = 2.3 \times 10^5\ r_g^2$, whereas for the pulsar B in the PSR J0737-3039 system the bound is about $10^4$ times larger \cite{NafJetzer}. On the other hand, in the strong-gravity regime \footnote{As opposed to the weak-gravity regime, where the chameleon effect \cite{Khoury_2004} takes over and the gravitational force is not modified.} (for a variety of equations of state by using observational constraints on the mass-radius relation), the constraint is $\alpha \lesssim 10^{10}\  \text{cm}^2 = 0.5\ r_g^2$ \cite{Arapoglu}. In our work, the condition of stability imposes a new constraint given by $\alpha \lesssim 0.01\ r_g^2$.

From Eqs. (\ref{2})-(\ref{4}) together with the four-divergence of the energy-momentum tensor, within the framework of quadratic gravity, the structure of a star in the state of hydrostatic equilibrium is described by \textit{modified Tolman-Oppenheimer-Volkoff (TOV) equations} \cite{1512.05711}:
\begin{subequations}
\begin{align}
    \frac{d\psi}{dr} &= \dfrac{1}{4r( 1+ 2\alpha R + \alpha rR' )}\bigg[ r^2e^{2\lambda}(2\kappa p^{\text{m}} - \alpha R^2) + 2(1 + 2\alpha R)\left( e^{2\lambda} -1 \right) - 8\alpha r R' \bigg] , \label{5a}  \\
%
    \frac{d\lambda}{dr} &= \dfrac{1}{4r(1+ 2\alpha R + \alpha rR')}\left\lbrace 2(1 + 2\alpha R)\left(1-e^{2\lambda}\right)  + \dfrac{r^2e^{2\lambda}}{3}\bigg[ 2\kappa(2\epsilon^{\text{m}} + 3p^{\text{m}}) + 2R + 3\alpha R^2  \bigg]\right.  \nonumber  \\
    & \hspace{0.05\textwidth}\left. + \dfrac{2\alpha rR'}{1+2\alpha R}\left[ 2(1 + 2\alpha R)\left(1-e^{2\lambda}\right) + \dfrac{r^2e^{2\lambda}}{3}(2\kappa\epsilon^{\text{m}} + R + 3\alpha R^2) + 4\alpha rR' \right]\right\rbrace , \label{5b}   \\  
%
    \frac{d^2R}{dr^2} &= \dfrac{e^{2\lambda}}{6\alpha}\bigg[ R + \kappa\big(3p^{\text{m}} - \epsilon^{\text{m}}\big) \bigg] + \left( \lambda' - \psi' - \dfrac{2}{r} \right)R' ,  \label{5c}  \\
    \frac{dp^{\text{m}}}{dr} &= -(\epsilon^{\text{m}} + p^{\text{m}})\psi' ,   \label{5d}
\end{align}
\end{subequations}
where the prime denotes derivative with respect to the radial coordinate. As usual, the radius of the star $r_{\rm surf}$ is determined by the condition $p^{\text{m}}(r_{\rm surf}) = 0$.

In order to close the system of four coupled differential equations (\ref{5a})-(\ref{5d}) (equivalent to five $1^{st}$-order ones), we need an EoS which relates the pressure and the density of ordinary matter inside the star, this is, $p^{\text{m}} = p^{\text{m}}(\epsilon^{\text{m}})$. To guarantee regularity at the origin, we should establish the following five boundary conditions:
\begin{align}\label{6}
    \psi(0) &= 0,   &   \lambda(0) &= 0,   &    R(0) &= R_{\rm cent},   &    R'(0) &= 0 , \ 
\end{align}
for a given central energy density $\epsilon^{\text{m}}(0) =\epsilon^{\text{m}}_{\rm cent}$. Outside the star (where both density and pressure of the ordinary fluid vanish), we have to solve Eqs. (\ref{5a})-(\ref{5c}) --- equivalent to four $1^{st}$-order ones --- subject to the four junction conditions on the stellar surface
\begin{align}\label{7}
\psi_{in}(r_{\rm surf}) &= \psi_{out}(r_{\rm surf}),   &   \lambda_{in}(r_{\rm surf}) &= \lambda_{out}(r_{\rm surf}),   \nonumber   \\
R_{in}(r_{\rm surf}) &= R_{out}(r_{\rm surf}),    &   R'_{in}(r_{\rm surf}) &= R'_{out}(r_{\rm surf}).
\end{align}
The value of Ricci scalar at the center ($R_{\rm cent}$) must be chosen so that it satisfies the asymptotic flatness requirement at infinity, namely $R(r) \rightarrow 0$ as $r \rightarrow \infty$.


\section{Curvature fluid}\label{Sec3}

We describe the curvature-induced terms as an effective fluid. In other words, field equations in $f(R)$ gravity (\ref{2}) can be written as effective Einstein equations with a total energy-momentum tensor composed of two fluids indeed: a \textit{standard matter fluid}, described by $T_{\mu\nu}^{\text{m}}$, and a \textit{curvature fluid}, described by,
\begin{equation}\label{8}
T_{\mu\nu}^{\text{c}} \equiv  \dfrac{1}{\kappa}\bigg[ (1 - f_R)R_{\mu\nu} + \dfrac{1}{2}(f -R)g_{\mu\nu} +  \nabla_\mu\nabla_\nu f_R - g_{\mu\nu}\square f_R \bigg] ,
\end{equation}
and accordingly, the Einstein tensor is given by
\begin{equation}\label{EFE}
G_{\mu\nu} = \kappa\left( T_{\mu\nu}^{\text{m}} + T_{\mu\nu}^{\text{c}} \right) .
\end{equation}
Such approach has already been successfully used in the study of cosmological perturbations \cite{Carloni, CapozzielloLauLam}. To the best of our knowledge, the present paper is the first time such definitions are used to investigate the (radial) perturbations of a star.

Eq. (\ref{8}) can be put in the form of an energy-momentum tensor corresponding to an anisotropic perfect fluid, i.e., $T_\mu^{(\text{c})\nu} = \text{diag}\left( -\epsilon^{\text{c}}, p^{\text{c}}, p^{\text{c}}_t, p^{\text{c}}_t \right)$. In particular, for a stellar configuration  in  hydrostatic equilibrium described by the metric (\ref{4}) in the Starobinsky model, the curvature energy density, the curvature radial pressure, and the curvature transverse pressure are given, respectively, by 
\begin{align}
    \epsilon^{\text{c}} &\equiv \dfrac{1}{\kappa}\left\lbrace -\dfrac{2\alpha R}{r^2}\dfrac{d}{dr}\bigg[ r\left( 1- e^{-2\lambda} \right)\bigg] + \dfrac{\alpha}{2}R^2 + \dfrac{2\alpha}{e^{2\lambda}}\left[ \left( \dfrac{2}{r}- \lambda' \right)R' + R'' \right] \right\rbrace ,  \label{9}  \\
p^{\text{c}} &\equiv \dfrac{1}{\kappa}\left\lbrace -\dfrac{2\alpha R}{r}\left[ \dfrac{2\psi'}{e^{2\lambda}} - \dfrac{1}{r}\left( 1-e^{-2\lambda} \right) \right] - \dfrac{\alpha}{2}R^2 - \dfrac{2\alpha}{e^{2\lambda}}\left( \dfrac{2}{r} + \psi' \right)R'  \right\rbrace ,   \label{10}  \\
p^{\text{c}}_t & \equiv \dfrac{1}{\kappa}\left\lbrace -\dfrac{2\alpha R}{r^2} \bigg[ 1 + \frac{r\lambda' - r\psi' -1}{e^{2\lambda}} \bigg] + \dfrac{\alpha}{2}R^2 - \dfrac{2\alpha}{e^{2\lambda}}\left[ \left( \dfrac{1}{r}+ \psi' - \lambda' \right)R' + R'' \right] \right\rbrace .  \label{11}
\end{align}

From now on, we choose to disregard the curvature transverse pressure $p^{\text{c}}_t$ since there are no growing modes for this quantity \cite{Tatekawa:2004mq}. Nevertheless, we point out that, even in GR, a non-vanishing transverse pressure at the star surface (as in the present case) modifies the value of the central density above which stability ceases to exist \cite{Arba_il_2016}. Here, we do expect such change --- we recall that we write the modified equations as GR with an extra ``effective'' (curvature) component --- with, perhaps, a different sign and magnitude of the change. Indeed, as we will show in  Section~\ref{Sec5}, the standard stable branch turns out to be unstable for not-so-small values of $\alpha$.

Once the system of equations (\ref{5a})-(\ref{5d}) is solved for a given EoS, the curvature quantities are obtained for each specific value of central density. 

The $tt$ component of the field equations (\ref{2}) leads to an analogous expression in GR:
\begin{equation}\label{12}
    \frac{d}{dr}\big(re^{-2\lambda}\big) - 1 = - \kappa r^2(\epsilon^{\text{m}} + \epsilon^{\text{c}}) \equiv - \kappa r^2\epsilon ,
\end{equation}
which allows us to introduce a mass function $m(r)$ through the relation
\begin{equation}\label{13}
    e^{-2\lambda} = 1 - \frac{2Gm(r)}{c^2r} ,
\end{equation}
or alternatively
\begin{equation}\label{mass}
\frac{dm}{dr} = \frac{4\pi}{c^2} r^2\epsilon^{\rm m} + \frac{\alpha c^2r^2}{2G}  \left\lbrace -\dfrac{2R}{r^2}\dfrac{d}{dr}\bigg[ r\left( 1- e^{-2\lambda} \right)\bigg] + \dfrac{R^2}{2} + \dfrac{2}{e^{2\lambda}}\left[ \left( \dfrac{2}{r}- \lambda' \right)R' + R'' \right] \right\rbrace . 
\end{equation}
The second term on the right-hand side of Eq. (\ref{mass}) plays a crucial role when we determine the mass of a compact star in the Starobinsky model. This means that within the star both the ordinary matter fluid and the curvature fluid contribute to the mass enclosed by a sphere of radius $r \leq r_{\rm surf}$. Meanwhile outside the star, unlike GR where $R = 0$, there is an energy density due to the curvature fluid described by $R \neq 0$. As a result, the star is surrounded by a sphere (dubbed  ``gravitational sphere'' \cite{AstashenokCapOdi3, AstashenokOdinDom}) that contributes to the gravitational mass \cite{Fulvio}. In fact, there is an effective radius $r_{\text{eff}}$ beyond which the mass function becomes a constant (up to $0.01\%$) \footnote{there is no observable variation of the frequency of the fundamental mode with a variation of $0.1\%$}. We take this radius to be the effective boundary of the star and henceforth we are going to define a total pressure as $p \equiv p^{\text{m}} + p^{\text{c}}$.


\section{Equations for radial oscillations} \label{Sec4} 

The study of radial adiabatic oscillations (for which the assumption of spherical symmetry holds and the dissipative flows play no role) within the framework of GR began with Chandrasekhar \cite{Chandrasekhar1, Chandrasekhar}, where Einstein equations are linearized around the equilibrium configuration to generate a Sturm-Liouville problem. Equations that govern such pulsations can be rewritten in several forms \cite{Chanmugam, KokkotasRuoff, VathChanmugam, Gondek, PanotopoulosLopes2017}, some of them being suitable for numerical computations. Here, we will follow Gondek's form \cite{Gondek}, namely, a system of coupled first-order time-independent equations given by
\begin{align}\label{14}
    \dfrac{d\zeta}{dr} &= -\dfrac{1}{r}\left( 3\zeta + \dfrac{\Delta p}{\gamma p} \right) - \dfrac{dp}{dr}\dfrac{\zeta}{p+ \epsilon} , \\
\label{15}
    \dfrac{d(\Delta p)}{dr}  &= \zeta\left[ \dfrac{\omega^2}{c^2}e^{2(\lambda - \psi)}(p + \epsilon)r - 4\dfrac{dp}{dr} - \kappa e^{2\lambda}(p + \epsilon)r p + \dfrac{r}{p + \epsilon}\left( \dfrac{dp}{dr} \right)^2  \right] +  \nonumber  \\
    & \quad + \Delta p\left[ \dfrac{1}{p+ \epsilon}\dfrac{dp}{dr} - \frac{\kappa}{2} (p + \epsilon)r e^{2\lambda} \right] ,
\end{align}
where $\xi(t,r) \equiv \chi(r)e^{i\omega t}$ is the Lagrangian displacement, $\zeta \equiv \chi/r$, $\gamma \equiv (1 + \epsilon/p)dp/d\epsilon$ is the \textit{adiabatic index},  $\Delta p \equiv \delta p + \chi \, dp/dr$ is the Lagrangian perturbation of the pressure, and $\delta p$ is the Eulerian perturbation.  

In order to guarantee that $d\zeta/dr$ is regular everywhere in Eq. (\ref{14}), we require that $\Delta p = -3\gamma\zeta p$ as $r\rightarrow 0$. In GR the other boundary condition is imposed on the surface of the star where the matter pressure is zero. In the present work, however, we have at the surface $p \equiv p^{\text{m}} + p^{\text{c}} \neq 0$ because of $p^{\text{c}}(r_{\rm surf}) \neq 0$. Thus, it is convenient to establish the boundary condition at $r_{\text{eff}}$ where the {\it total} pressure goes to zero, so that the Lagrangian perturbation of the pressure must satisfy $\Delta p = 0$ as $r\rightarrow r_{\text{eff}}$.


\section{Numerical results and discussion}\label{Sec5}

\begin{figure}[t]
\centering 
\includegraphics[width=0.60\textwidth,origin=c]{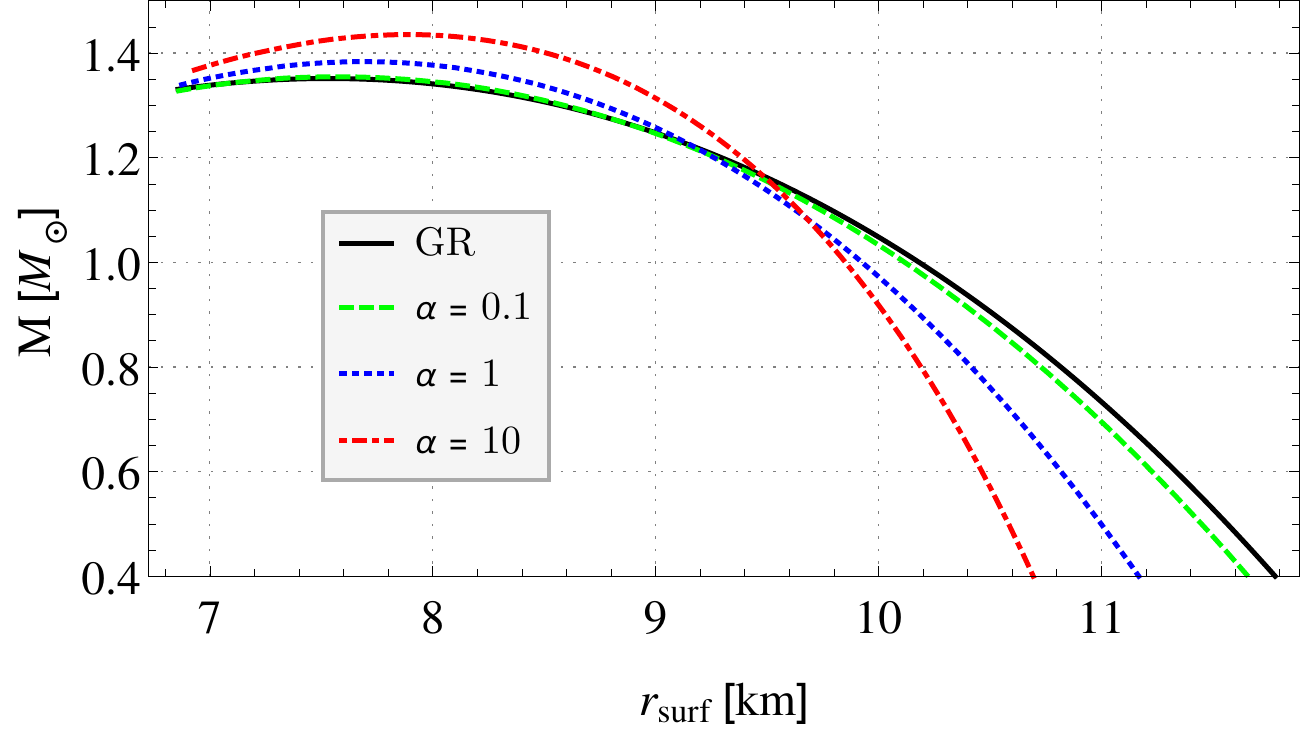} 
\includegraphics[width=0.60\textwidth,origin=c]{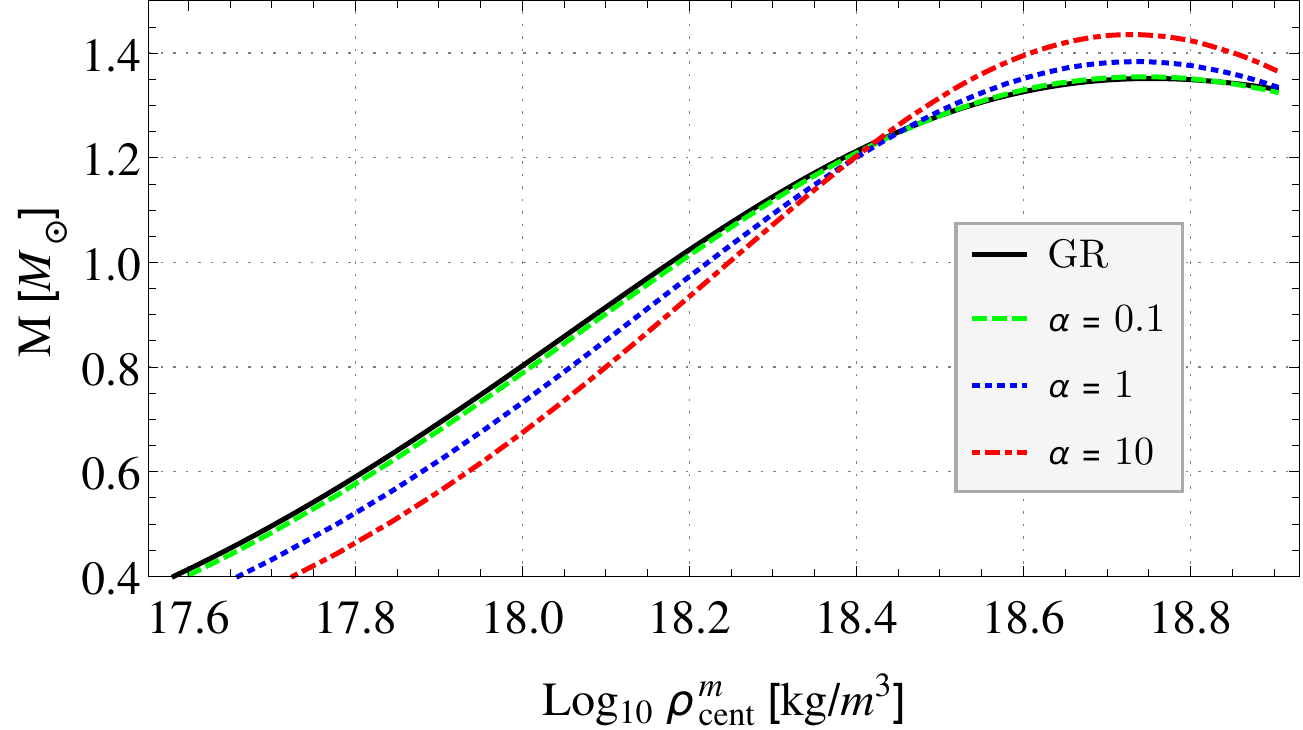}
\caption{\label{figure1} Stellar parameters for neutron stars with polytropic  EoS in GR and in $f(R) = R + \alpha R^2$ gravity for some values of the free parameter $\alpha$ in $r_g^2$ units.  Upper panel: Mass-radius relation. Lower panel: Mass-central density diagram. As it has been shown in previous works, the maximum mass increases as the value of $\alpha$ increases. }
\end{figure}

The first stage of our work is to numerically solve the modified TOV equations (\ref{5a})-(\ref{5d}) inside and outside the star with initial and boundary conditions (\ref{6}) and (\ref{7}) for a given central density $\epsilon^{\text{m}}_{\rm cent}$. This is performed taking into account that the asymptotic-flatness requirement is fulfilled only for a unique value of $R_{\rm cent}$. Each solution provides us a configuration in state of hydrostatic equilibrium with surface radius $r_{\rm surf}$ and total gravitational mass $M = m(r_{\text{eff}})$. Families of equilibrium solutions, in GR and in the Starobinsky model for three values of $\alpha$, are presented in Fig. \ref{figure1}; the upper panel shows the mass-radius curves and the lower panel shows the mass-central density relations. 

According to GR, stable stars have radii (central densities) larger (smaller) than the value corresponding to the maximum mass in the mass-radius diagram. In $f(R)$, however, it is crucial to realize that even smaller-mass solutions could be in a state of either stable or unstable equilibrium. In other words, this equilibrium configurations pictured in Fig.~\ref{figure1} does not guarantee stability with respect to a compression or decompression generated by a radial perturbation. Indeed, in GR \cite{Glendenning} the transition from stability to instability always occurs for isotropic stars at the first maximum on the mass-central density curve. Nevertheless,  this condition is no longer satisfied in $f(R)$ gravity because of the contribution from the curvature fluid. Namely, there are two main reasons for such behavior: First, a fluid with anisotropic stress shifts \cite{Arba_il_2016} the stability threshold value for the central density, which no longer coincides with the corresponding value for the maximum mass, even in GR. This is precisely the case for the curvature fluid at hand.  Secondly, the curvature fluid ``leaks out'' from the ``barionic" star itself (whose surface radius is defined by $p^{\text{m}}(r_{\rm surf}) = 0$), but the effective radius, which defines the mass, is larger than that.

Therefore, the sufficient condition for stability, that holds in the most general cases and which we shall adopt here, is that the frequencies of the normal modes of radial oscillations must be Real. If the squared frequency of the fundamental mode is negative for a particular configuration, then at least this one frequency is imaginary and, therefore, the star is unstable, since the perturbation will then follow the growing mode $\sim \exp(+ \Im(\omega) t )$ --- which is absent from the beginning in quasinormal modes studies \cite{Kokkotas1999}.

The numerical integration of Eqs. (\ref{14}) and (\ref{15}) is carried out by using the shooting method, that is, we integrate the equations for a set of trial values of $\omega^2$ satisfying the condition  $\Delta p = -3\gamma\zeta p$ at $r = 0$. In addition, we consider normalized eigenfunctions  $(\zeta(0) = 1)$ at the center, and we integrate up to the effective radius. Then, the values of the squared frequency for which the boundary condition $\Delta p (r_{\text{eff}}) = 0$ is satisfied are the correct normal frequencies of the radial pulsations. 

\begin{figure}[t]
\centering 
\includegraphics[width=0.60\textwidth,origin=c]{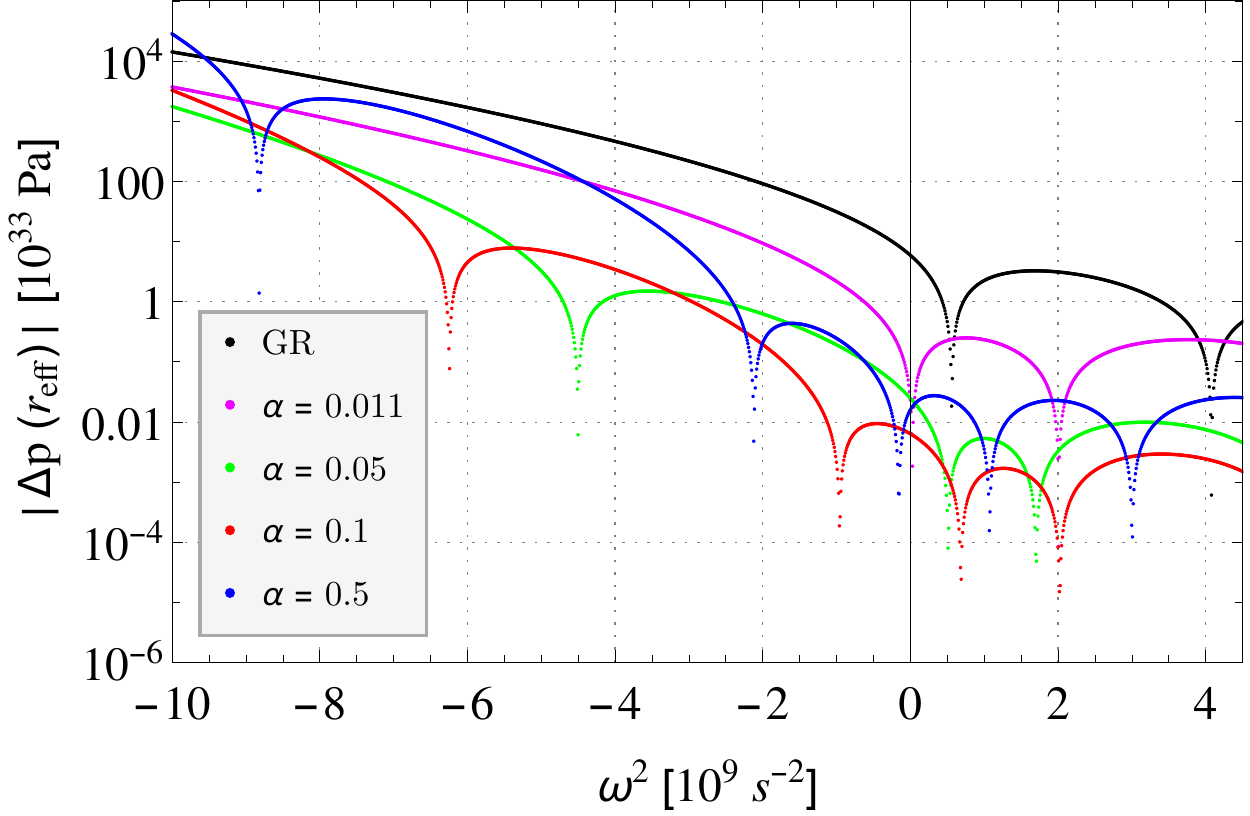} 
\caption{\label{figure2} Absolute value of the Lagrangian perturbation of the pressure at the effective radius on a logarithmic scale for a set of trial values of $\omega^2$ in GR and Starobinsky model with a central mass density $\rho_{\rm cent}^{\text{m}} = 2.0 \times 10^{18}\ \text{kg}/\text{m}^3$. The minima in each curve correspond to the correct frequencies of the oscillation modes for equilibrium configurations. The values of the parameter $\alpha$ (see inset) are given in units of $r_g^2$.}
\end{figure}

In particular, we show in Fig.~\ref{figure2} the Lagrangian perturbation of the pressure at $r_{\text{eff}}$ for a set of values $\omega^2$ in GR and in $f(R) = R + \alpha R^2$ gravity, for a central mass density $\rho_{\rm cent}^{\text{m}} = 2.0 \times 10^{18}\ \text{kg}/\text{m}^3$. Such a star is stable according to GR (see Fig.~\ref{figure1}, lower panel).
The first (leftmost) minimum in each curve corresponds to the frequency of the fundamental mode. If this mode is stable ($\omega_0^2 > 0$), then all radial modes are stable since the fundamental frequency is the lowest one.
As a consequence of this stability requirement, for the central density considered, the free parameter must be $\alpha \lesssim 0.011\ r_g^2 = 2.4 \times 10^8\ \text{cm}^2$. Such tiny value would render the quadratic gravity indistinguishable from GR for all previously proposed tests in the literature. In particular, we point out that such small values of $\alpha$ would yield minor and unobservable changes in the maximum mass (note the order of magnitude of the values for $\alpha$ used in Fig.~\ref{figure1}).
We have tested the procedure for two other values of central density in the GR stable branch and obtained compatible results.

\section{Conclusion}\label{Sec6}

The results presented here are stronger than previous values in the literature. We have pointed out possible reasons for such discrepancies; namely, the curvature fluid itself --- in spite of previous success in Cosmology --- its anisotropic stress and the subsequent different radius definitions. The Ricci scalar, which is an extra degree of freedom in $f(R)$ theories, is implicitly embedded in the effective fluid and, therefore, this could lead to different results when compared to works that explicitly introduce a perturbation on this quantity, as in Ref. \cite{Folomeev2020}. In particular, we have made two assumptions: First, we have neglected the tangential component of the pressure corresponding to the curvature fluid (which is always present and under no limit becomes isotropic) and, second, the equations for radial oscillations were not integrated up to the stellar surface (where only the matter pressure is zero) but up to the effective radius where the total radial pressure vanishes and the boundary condition is satisfied. In addition, we have also preformed a full-equation analysis (i.e, non perturbative) that yields more robust results. We will keep following such analysis of the modified Einstein equations both with and without the curvature-fluid definition for different models in a series of forthcoming works.

We are currently exploring more realistic EoSs as well as viable modified gravity models. Furthermore, we are investigating perturbations of the dynamical equation for the Ricci scalar and the field equations in order to verify the results obtained here.

\acknowledgments

JMZP acknowledges Brazilian funding agency CAPES for PhD scholarship 331080/2019.


\end{document}